\documentclass{article}

\usepackage{microtype}
\usepackage{graphicx}
\usepackage{subcaption}
\usepackage{booktabs}

\usepackage{hyperref}

\usepackage[accepted]{sysml2019}

\begin{document}

\twocolumn[
\sysmltitle{SqueezeWave: Extremely Lightweight Vocoders for On-device Speech Synthesis}

\sysmlsetsymbol{equal}{*}

\begin{sysmlauthorlist}
\sysmlauthor{Bohan Zhai}{equal,to}
\sysmlauthor{Tianren Gao}{equal,to}
\sysmlauthor{Flora Xue}{to}
\sysmlauthor{Daniel Rothchild}{to}\\
\sysmlauthor{Bichen Wu}{to}
\sysmlauthor{Joseph E. Gonzalez}{to}
\sysmlauthor{Kurt Keutzer}{to}
\end{sysmlauthorlist}

\sysmlaffiliation{to}{Department of EECS, UC Berkeley}
\sysmlcorrespondingauthor{Bichen Wu}{bichen@berkeley.edu}

\sysmlkeywords{Machine Learning, SysML}

\vskip 0.3in

\begin{abstract}
Automatic speech synthesis is a challenging task that is becoming increasingly important as edge devices begin to interact with users through speech. Typical text-to-speech pipelines include a vocoder, which translates intermediate audio representations into an audio waveform. Most existing vocoders are difficult to parallelize since each generated sample is conditioned on previous samples.  WaveGlow is a flow-based feed-forward alternative to these auto-regressive models \cite{waveglow}. However, while WaveGlow can be easily parallelized, the model is too expensive for real-time speech synthesis on the edge. This paper presents SqueezeWave, a family of lightweight vocoders based on WaveGlow that can generate audio of similar quality to WaveGlow with 61x - 214x fewer MACs. Code, trained models, and generated audio are publicly available at \url{https://github.com/tianrengao/SqueezeWave}. 
\end{abstract}
]

\printAffiliationsAndNotice{\sysmlEqualContribution} 

\section{Introduction}

Edge devices are increasingly interacting with users through speech: map applications read out directions to drivers; home assistant devices engage in natural language exchanges with users; translation apps speak text translated into a foreign language. Applications like these rely on automatic text-to-speech (TTS) algorithms, which in recent years have approached true speech in sound quality \cite{waveglow, wavenet, deepvoice}. Driving these advances in TTS sound quality are complex machine learning models, which require significant computing resources to run. As a result, to satisfy the latency constraints of many applications, it is only feasible to run these models in the cloud, and send the synthesized audio to edge devices.

However, a number of trends are challenging this paradigm.
First, hardware used in mobile phones is becoming increasingly more powerful, and making effective use of this computation could lead to significant reductions in cloud computing costs.
Second, consumers are becoming increasingly concerned about data privacy, especially
concerning speech data.
Smartphones, smart TVs, and home assistants have all been accused of sending sensitive data to the cloud without users' knowledge\footnote{https://www.washingtonpost.com/technology/2019/05/06/alexa-has-been-eavesdropping-you-this-whole-time} . Moving the machine learning computations to the edge would eliminate the need to send data to the cloud in the first place.
Finally, consumers are becoming increasingly reliant on speech synthesis systems, to provide timely driving directions, to respond interactively to questions, etc. These applications must work with low latency and even without a reliable Internet connection -- constraints that can only be satisfied when speech synthesis is done on-device. Responding to these trends requires moving the TTS models to the edge.

Modern TTS systems typically consist of two steps: a synthesizer first generates acoustic features (e.g., a mel-spectrogram) from text inputs, and a vocoder then generates waveforms from those acoustic features. This paper focuses on improving the efficiency of vocoders. Existing vocoders produce high-quality speech at high computational cost. Existing vocoders, such as WaveNet \cite{wavenet} and its variants \cite{WaveRNN,lpcnet}, are auto-regressive, meaning that each generated sample depends on previous samples. 
The computation of auto-regressive models are inherently serial, and the fine-grained sequential nature of this computation hinders parallel hardware acceleration, increasing inference latency and making real-time deployment infeasible. Recently, Pregner et al. proposed WaveGlow~\cite{waveglow}, a flow-based speech generation model. WaveGlow is not auto-regressive, and the model generates many samples in each forward pass. Although this makes WaveGlow highly parallelizable, the model is computationally expensive, requiring 229G MACs to generate 1 second of 22kHz speech -- far beyond the capability of mobile processors. As a result, although WaveGlow can reach faster-than-realtime speed on the latest NVIDIA V100 GPUs, it is not suitable for edge deployment.

In this paper, we propose SqueezeWave, a family of extremely lightweight flow-based vocoders for on-device speech synthesis. Previous work \cite{iandola2016squeezenet, SqueezeDet, wu2017shift, wu2018squeezeseg, wu2018squeezesegv2, yang2018synetgy, gholami2018squeezenext,Wu:EECS-2019-120,wu2019fbnet,Dai_2019_CVPR} have shown that optimizing the neural network architecture can lead to significant efficiency improvement in many applications.  Therefore, in this work,  we carefully re-design Waveglow's network architecture. By re-arranging the audio tensor, adopting depthwise separable convolutions, and making other optimizations, we reach SqueezeWave, a family of vocoders that can generate high-quality speech with 61-214x fewer MACS than WaveGlow. On a Macbook Pro with an Intel i7 CPU, SqueezeWave generates waveforms at a speed of 123K - 303K samples per second, or 5.6x - 13.8x faster than the real-time. Even on a Raspberry Pi 3B+ with a Broadcom BCM2837 CPU, we are able to reach a near real-time speed of 15.6K samples per second. Our code, trained models, and generated samples are publicly available at \url{https://github.com/tianrengao/SqueezeWave}.

\section{Computational complexity of WaveGlow}
WaveGlow is a flow-based model that generates an audio waveform conditioned on a mel-spectrogram. WaveGlow consists of a sequence of bijections that progressively transform a waveform into a latent space. The bijections are conditioned on the text, and they are trained to transform the data distribution into a Gaussian distribution in the latent space. During inference, the model draws a Gaussian sample and transforms it back to the data distribution.

Instead of convolving the waveforms directly, WaveGlow first groups nearby samples to form a multi-channel input $x \in R^{L,C_g}$, where $L$ is the length of the temporal dimension and $C_g$ is the number of grouped audio samples per time step (The number of samples in the waveform is just $L \times C_g$). This grouped waveform $\mathbf{x}$ is then transformed by a series of bijections, each of which takes $\mathbf{x}^{(i)}$ as input and produces $\mathbf{x}^{(i+1)}$ as output. Within each bijection, the input signal $\mathbf{x}^{(i)}$ is first processed by an invertible point-wise convolution, and the result is split along the channel dimension into $\mathbf{x}_a^{(i)}, \mathbf{x}_b^{(i)} \in \mathbf{R}^{L, C_g/2}$. $\mathbf{x}_a^{(i)}$ is then used to compute affine coupling coefficients $(\log \mathbf{s}^{(i)}, \mathbf{t}^{(i)}) = WN(\mathbf{x}_a^{(i)}, \mathbf{m})$. $\mathbf{s}^{(i)}, \mathbf{t}^{(i)} \in \mathbf{R}^{L, C_g/2}$ are the  affine coupling coefficients that will be applied to $\mathbf{x}_b^{(i)}$, $WN(\cdot, \cdot)$ is a WaveNet-like function, or WN function for short, $\mathbf{m} \in \mathbf{R}^{L_m, C_m}$ is the mel-spectrogram that encodes the audio, $L_m$ is the temporal length of the mel-spectrogram and $C_m$ is the number of frequency components. Next, the affine coupling layer is applied: $\mathbf{x}_b^{(i+1)} = \mathbf{x}_b^{(i)} \bigotimes \mathbf{s}^{(i)}  + \mathbf{t}^{(i)}, \mathbf{x}_a^{(i+1)} = \mathbf{x}_a^{(i)}$, where $ \bigotimes$ denotes element-wise multiplication. Finally, $\mathbf{x}_a^{(i)}$ and $\mathbf{x}_b^{(i+1)}$ are concatenated along the channel dimension. 

\begin{figure*}[h]
\begin{subfigure}[t]{.48\textwidth}
    \centering
    \includegraphics[width=\linewidth]{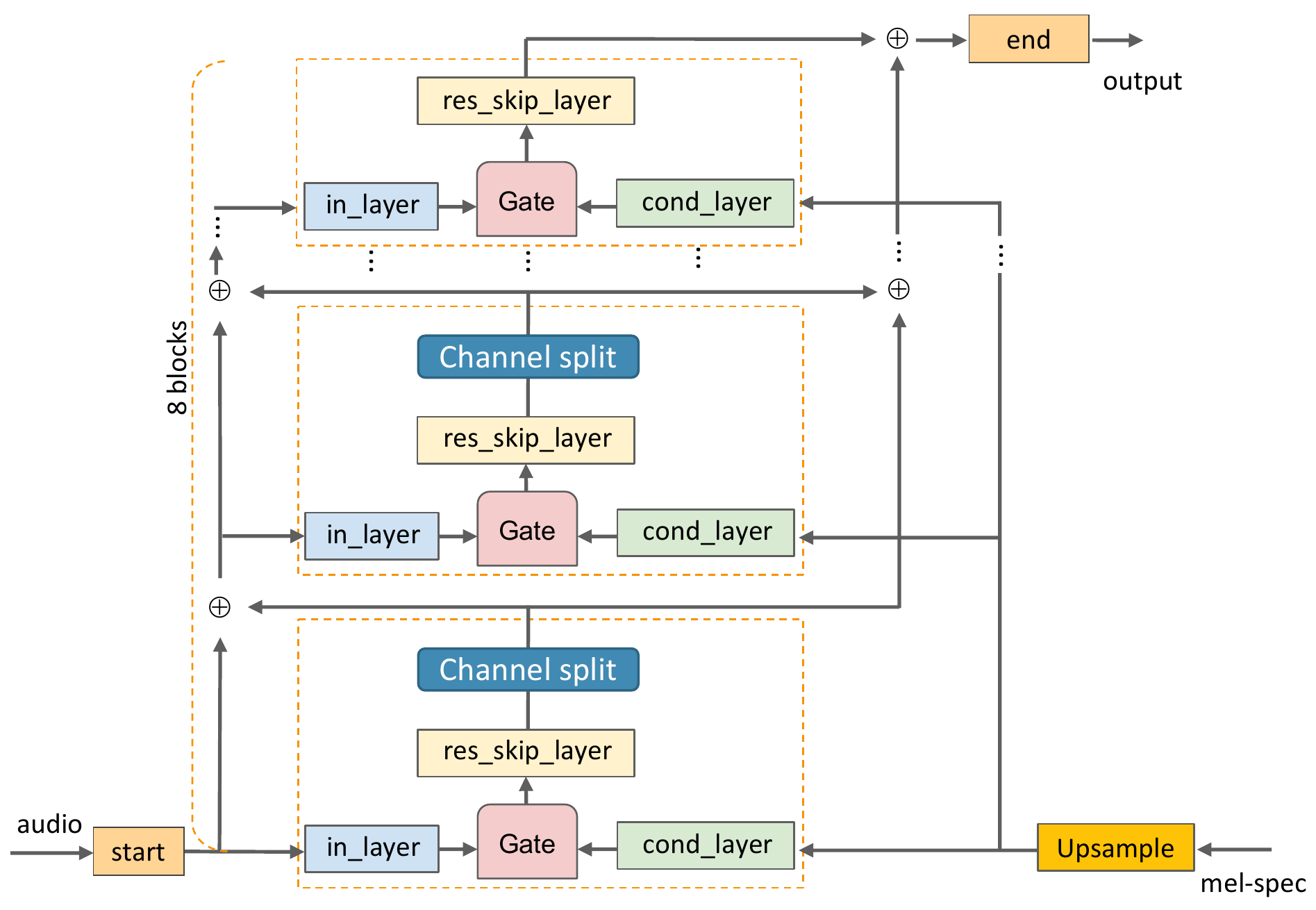}
    \caption{Structure of the WN function in WaveGlow.}
    \label{fig:WaveGlow-WN}
\end{subfigure}
\begin{subfigure}[t]{.48\textwidth}
    \centering
    \includegraphics[width=\linewidth]{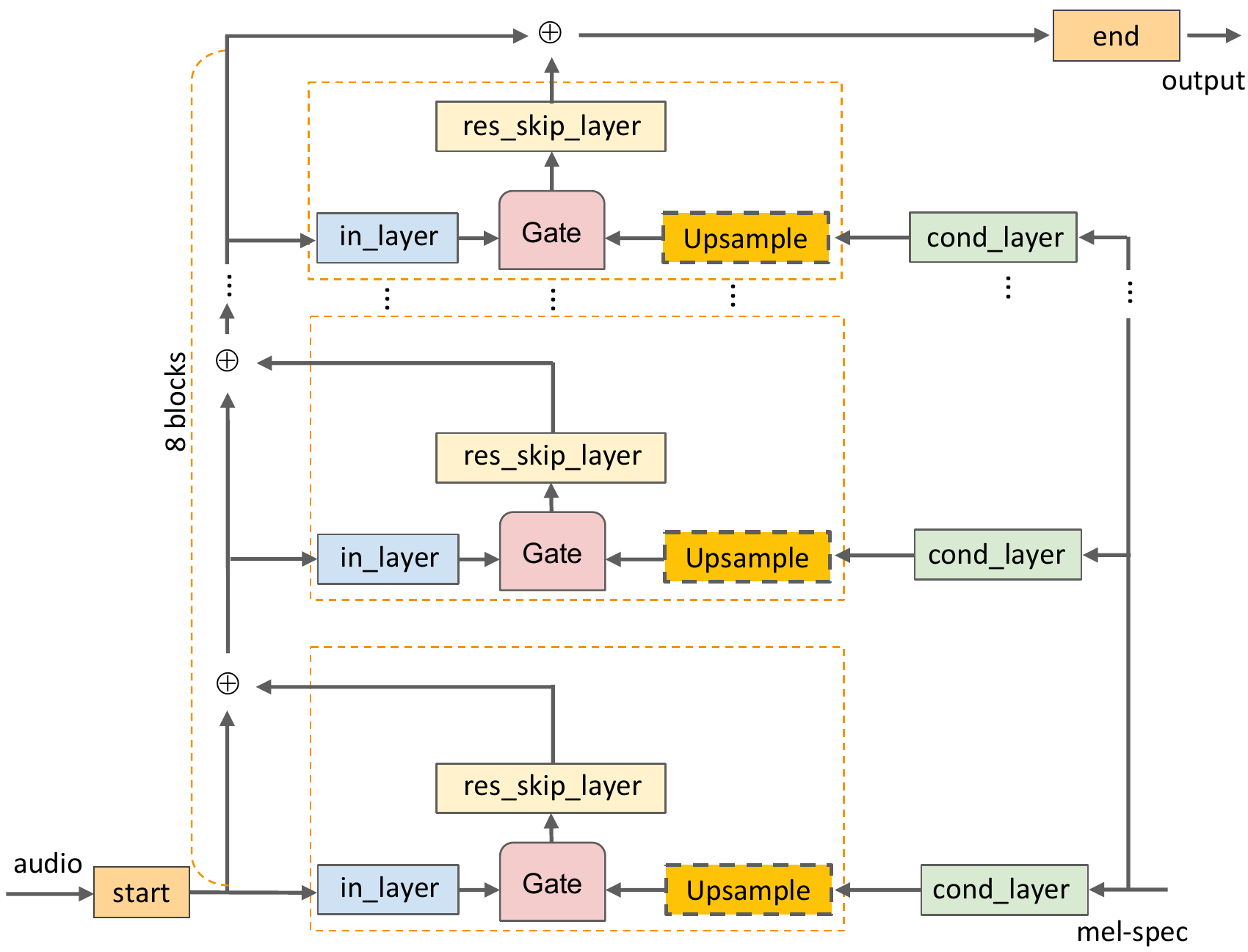}
    \caption{Structure of the WN function in SqueezeWave.}
    \label{fig:SqueezeWave-WN}
\end{subfigure}
\end{figure*}

The majority of the computation of WaveGlow is in the WN functions $WN(\cdot, \cdot)$, illustrated in Figure \ref{fig:WaveGlow-WN}. The first input to the function is processed by a point-wise convolution labeled \textit{start}. This convolution increases the number of channels of $\mathbf{x}_a^{(i)}$ from $C_g/2$ to a much larger number. In WaveGlow, $C_g=8$, and the output channel size of \textit{start} is 256.  Next, the output is processed by a dilated 1D convolution with a kernel size of 3 named \textit{in\_layer}. Meanwhile, the mel-spectrogram $\mathbf{m}$ is also fed into the function. The temporal length of the mel-spectrogram $L_m$ is typically much smaller than the length of the reshaped audio waveform $L$. In WaveGlow, $L_m=63, C_m=80, L=2,000, C_g=8$. So in order to match the temporal dimension, WaveGlow upsamples $\mathbf{m}$, and then passes it through a convolution layer named \textit{cond\_layer}. The output of \textit{in\_layer} and \textit{cond\_layer} are combined in the same way as WaveNet \cite{wavenet} through the \textit{gate} function, whose output is then processed by a \textit{res\_skip\_layer}. The output of this layer has a temporal length of $L=2000$ and a channel size of 512 in the original WaveGlow. It is then split into two branches along the channel dimension. This structure is repeated 8 times and at the last one, the output of \textit{res\_skip\_layer} is then processed by a point-wise convolution named \textit{end}. This convolution computes the transformation factors $\mathbf{s}^{(i)}$ and $\mathbf{t}^{(i)}$ and compresses the channel size from 512 to $C_g=8$. 

According to the source code of WaveGlow, we calculate the computational cost of WaveGlow. To generate 1 second of 22kHZ audio, WaveGlow requires 229G MACs. Among all the layers, \textit{in\_layer}s accounts for 47\%, \textit{cond\_layer}s accounts for 39\%, and \textit{res\_skip\_layer} accounts for 14\%. The details of the calculation can be found in our source code.

\section{SqueezeWave}

\subsection{Reshaping audio waveforms}
After carefully examining the network structure of WaveGlow, we identified that a major source of the redundancy comes from the shape of the input audio waveform to the network. In the original WaveGlow, the input waveform is reshaped to have a large temporal dimension and small channel size ($L=2000, C_g=8$).  This leads to high computational complexity in three ways: 1) WaveGlow is a 1D convolutional neural network, and its computational complexity is linear in $L$. 2) Mel-spectrograms have a much coarser temporal resolution than the grouped audio: in the original WaveGlow, $L=2000$ but $L_m=63$. In order to match the temporal dimensions of the two signals, WaveGlow upsamples the mel-spectrogram before passing it through \textit{cond\_layer}s. The upsampled mel-spectrograms are highly redundant since new samples are simply interpolated from existing ones. Therefore, in WaveGlow, most of the computations in \textit{cond\_layer}s are not necessary. 3) Inside each WN function,  the 8-channel input is projected to have a large intermediate channel size, typically 256 or 512. A larger channel size is beneficial since it increases the model capacity. However, at the output of WN, the channel size is compressed to $C_g=8$ to match the audio shape. Such drastic reduction creates an ``information bottleneck'' in the network and information encoded in the intermediate representation can be lost. 

To fix this, we simply re-shape the input audio $\mathbf{x}$ to have a smaller temporal length and a larger channel size, while keeping the internal channel sizes within the WN function the same. In our experiments, we implement two settings: $L=64, C_g=256$ or $L=128, C_g=128$. (The total number of samples are changed from 16,000 to 16,384.) When $L=64$, the temporal length is the same as the mel-spectrogram, so no upsampling is needed. When $L=128$, we change the order of operators to first apply \textit{cond\_layer} on the the mel-spectrogram and then apply nearest-neighbor upsampling. This way, we can further reduce the computational cost of the \textit{cond\_layer}s.

\subsection{Depthwise convolutions}
Next, we replace 1D convolutions in the \textit{in\_layer} with depthwise separable convolutions. Depthwise separable convolutions are popularized by \cite{mobilenets} and are widely used in efficient computer vision models, including \cite{sandler2018mobilenetv2, wu2019fbnet}. In this work we adopt depthwise separable convolutions to process 1D audio.

\begin{figure}[h]
    \centering
    \includegraphics[width=\linewidth]{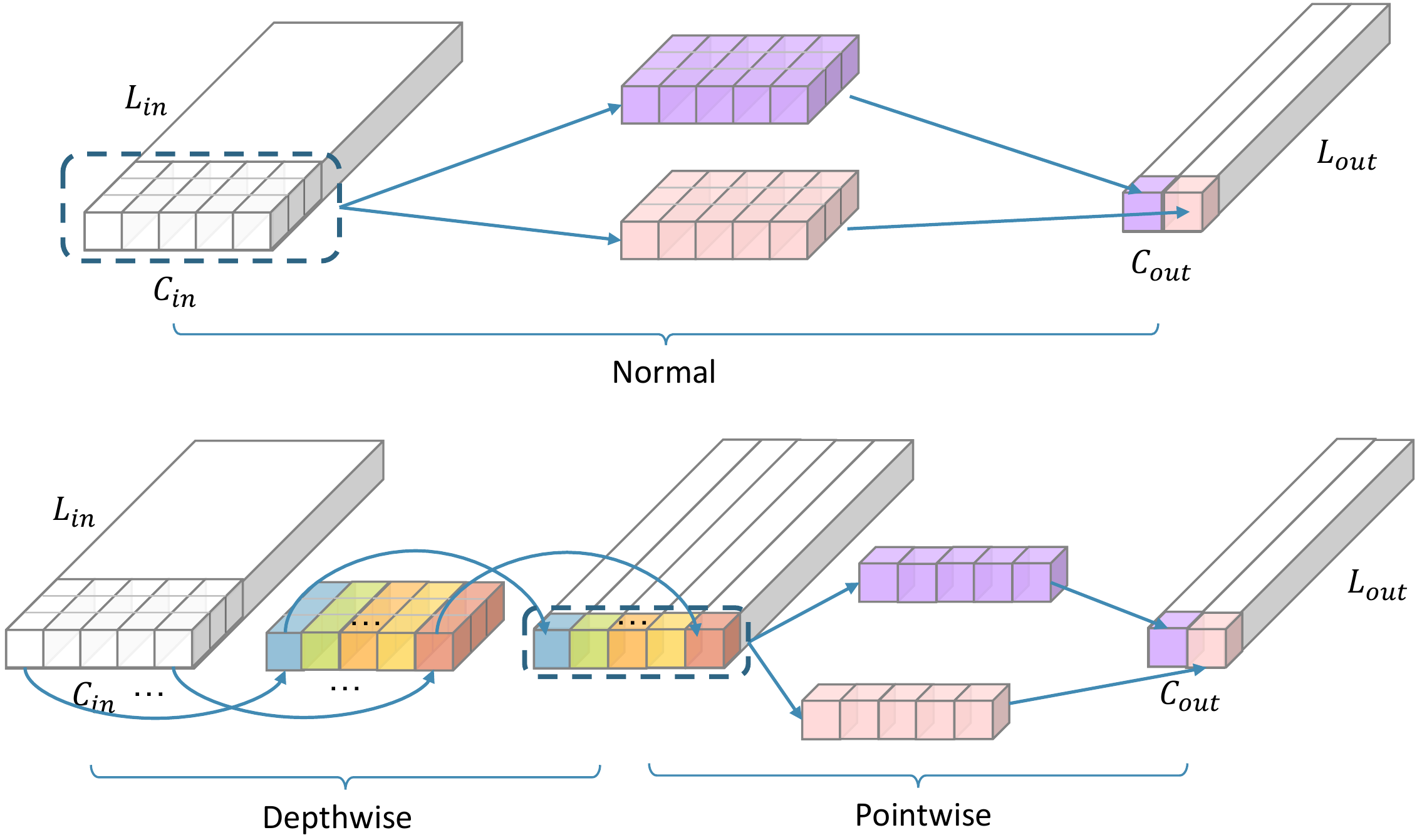}
    \caption{Normal convolutions vs. depthwise separable convolutions. Depthwise separable convolutions can be seen as a decomposed convolution that first combines information from the temporal dimension and then from the channel dimension.}
    \label{fig:dw}
\end{figure}

To illustrate the benefits of depthwise separable convolutions, consider a 1D convolutional layer that transforms an input with shape $C_{in}\times L_{in}$ into an output with shape $C_{out}\times L_{out}$, where $C$ and $L$ are the number of channels and temporal length of the signal, respectively.
For a kernel size $K$, the kernel has shape $K \times C_{in} \times C_{out}$, so the convolution costs $K \times C_{in} \times C_{out} \times L_{out}$ MACs. A normal 1D convolution combines information in the temporal and channel dimensions in one convolution with the kernel.
The depthwise separable convolution decomposes this functionality into two separate steps: (1) a temporal combining layer and (2) a channel-wise combining layer with a kernel of size 1.
Step 1 is called a depthwise convolution, and step 2 is called a pointwise convolution.
The difference between a normal 1D convolution and a 1D depthwise seperable convolution is illustrated in Figure \ref{fig:dw}.
After applying the depthwise separable convolution, the computational cost for step-1 becomes $K \times C_{in} \times L_{in}$ MACs
and for step-2, 
$C_{in} \times C_{out} \times L_{in}.$
The reduction of computation is therefore 
$$\frac{C_{in} \times C_{out} \times L_{in} + K \times C_{in} \times L_{in}}{K \times C_{in} \times C_{out} \times L_{in}} = \frac{1}{C_{out}} + \frac{1}{K}.$$  
In our setup, $K = 3$ and $C_{out} = 512$, so using this technique leads to around 3x MAC reduction in the $in\_layer$s.

\subsection{Other improvements}
In addition to the above two techniques, we also make several other improvements: 1) since the temporal length is now much smaller, WN functions no longer need to use dilated convolutions to increase the receptive fields, so we replace all the dilated convolutions with regular convolutions, which are more hardware friendly; 2) Figure \ref{fig:WaveGlow-WN} shows that the outputs of the \textit{res\_skip\_layer}s are split into two branches. Hypothesizing that such a split is not necessary since the topologies of the two branches are almost identical, we merge them into one and reduce the output channel size of the \textit{res\_skip\_layer}s by half. The improved SqueezeWave structure is illustrated in Figure \ref{fig:SqueezeWave-WN}. 


\section{Experiments}
In this section, we compare SqueezeWave with WaveGlow in terms of the efficiency and audio quality. We consider three metrics of computational efficiency: 1) MACs required per second of generated audio, 2) number of model parameters, and 3) actual speech generation speed, in generated samples per second, on a Macbook Pro and a Raspberry Pi 3b+. In terms of the audio quality, we use Mean Opinion Score (MOS) as the metric as in \cite{tacotron2, wavenet, waveglow, deepvoice}.

Our experimental setup is similar to that of \cite{waveglow}: we use the LJSpeech dataset \cite{ljspeech17}, which has 13,100 paired text/audio examples. We use a sampling rate of 22050Hz for the audio. We extract mel-spectrograms with librosa, using an FFT size of 1024, hop size 256 and window size 1024. We split the dataset into a training and a test set, and the split policy is provided in our source code. We reproduce the original WaveGlow model by training from scratch on 8 Nvidia V100 32GB RAM GPUs with a batch size 24. We train SqueezeWave with 24GB-RAM Titan RTX GPUs using a batch size of 96 for 600k iterations. Detailed configurations are available in our code. 

To evaluate the quality of the synthesized audio, we crowd-source our MOS evaluation on Amazon Mechanical Turk. We use 10 fixed sentences for each system, and each system/sentence pair is rated by 100 raters. Raters are not allowed to rate the same sentence twice, but they are allowed to rate another sentence from the same or a different system. We reject ratings that do not pass a hidden quality assurance test (ground truth vs. obviously unnatural audio). We report MOS scores with 95\% confidence intervals.

Tabel \ref{table:res} compares quality and efficiency of SqueezeWave and WaveGLow. WaveGlow achieves MOS scores comparable to those for ground-truth audio. However, the computational cost of WaveGlow is extremely high, as it requires 228.9 GMACs to synthesize 1 second of 22kHZ audio. SqueezeWave models are much more efficient. The largest model, SW-128L, with a configuration of L=128, $C_g$=256 requires 61x fewer MACs than WaveGlow. With reduced temporal length or channel size, SW-64L (106x fewer MACs) and SW-128S (214x fewer MACs) achieves slightly lower MOS scores but significantly lower MACs. Quantitatively, MOS scores of the SqueezeWave models are lower than WaveGlow, but qualitatively, their sound qualities are similar, except that audio generated by SqueezeWave contains some background noise. Noise cancelling techniques can be applied to improve the quality. Readers can find synthesized audio of all the models from our source code. We also train an extremely small model, SW-64S, with L=64, $C_g$=128. The model only requires 0.69 GMACs, which is 332x fewer than WaveGlow. However, the sound quality is obviously lower, as reflected in its MOS score. 

\begin{table}[h!]
    \centering
    \begin{tabular}{ccccc}
    \hline
         Models & MOS & GMACs & Ratio & Params \\
         \hline
         GT        & 4.62 $\pm$ 0.04 & -- & -- & -- \\
         WaveGlow  & 4.57 $\pm$ 0.04 & 228.9 & 1 & 87.7 M \\
         SW-128L   & 4.07 $\pm$ 0.06 & 3.78 & 61 & 23.6 M \\
         SW-128S   & 3.79 $\pm$ 0.05 &  1.07 & 214 & 7.1 M \\
         SW-64L    & 3.77 $\pm$ 0.05 & 2.16 & 106 & 24.6 M \\
         SW-64S    & 2.74 $\pm$ 0.04 & 0.69 & 332 & 8.8 M \\
    \hline
    \end{tabular}
    \caption{A comparison of SqueezeWave and WaveGlow. SW-128L has a configuration of L=128, $C_g$=256, SW-128S has L=128, $C_g$=128, SW-64L has L=64, $C_g$=256, and SW-64S has L=64, $C_g$=128. The quality is measured by mean opinion scores (MOS). The main efficiency metric is the MACs needed to synthesize 1 second of 22kHz audio. The MAC reduction ratio is reported in the column ``Ratio''. The number of parameters are also reported. 
    }
    \label{table:res}
\end{table}

\begin{table}[h!]
    \centering
    \begin{tabular}{cccc}
    \hline
         Models   & Macbook Pro & Raspberry Pi   \\
         \hline
         WaveGlow & 4.2K & Failed \\
         SW-128L  & 123K & 5.2K \\
         SW-128S  & 303K & 15.6K \\
         SW-64L   & 255K & 9.0K \\
         SW-64S   & 533K & 21K \\
    \hline
    \end{tabular}
    \caption{ Inference speeds (samples generated per second) on a Mackbook Pro and a Raspberry Pi. 
    }
    \label{table:speed}
\end{table}

We deploy WaveGlow and SqueezeWave to a Macbook Pro with an Intel i7 CPU and a Raspberry Pi 3B+ with a Broadcom BCM2837B0 CPU. We report the number of samples generated per second by each model in Table \ref{table:speed}. On a Mackbook, SqueezeWave can reach a sample rate of 123K-303K, 30-72x faster than WaveGlow, or 5.6-13.8x faster than real-time (22kHZ). On a Raspberry Pi computer, WaveGlow fails to run, but SqueezeWave can still reach 5.2k-21K samples per second. SW-128S in particular can reach near real-time speed while maintaining good quality. 

\bibliography{bib}
\bibliographystyle{sysml2019}

\end{document}